\documentclass[aps,prb,floatfix,twocolumn,superscriptaddress]{revtex4-2}
\usepackage{mathrsfs}
\usepackage{graphicx}
\usepackage{amsmath}
\usepackage{amssymb}
\usepackage{xcolor}

\usepackage{relsize}
\usepackage{textcomp}
\usepackage{dsfont}
\usepackage{bm}
\usepackage{float}

\newcommand{\me}{\mathrm{e}}      % base of the natural logarithm
\newcommand{\mi}{\mathrm{i}}      % imaginary unit

\newcommand{\dif}{\mathrm{d}}

\allowdisplaybreaks
\begin{document}

	\title{Metric-Changing Quasi-Hermitian Quenches: Normalized Loschmidt Amplitude and Complex-Time Zeros}
	
    \author{Xu-Yang Hou}
    \thanks{These authors contributed equally to this work.}
    \affiliation{School of Physics, Southeast University, Jiulonghu Campus, Nanjing 211189, China}

    \author{Jia-Chen Tang}
    \thanks{These authors contributed equally to this work.}
    \affiliation{School of Physics, Southeast University, Jiulonghu Campus, Nanjing 211189, China}
	
	\author{Hao Guo}
	\email{guohao.ph@seu.edu.cn}
	\affiliation{School of Physics, Southeast University, Jiulonghu Campus, Nanjing 211189, China}
	\affiliation{Hefei National Laboratory, University of Science and Technology of China, Hefei 230088, China}
	
\begin{abstract}
In quasi-Hermitian mixed-state quenches, the initial and final Hamiltonians generally possess different positive-definite metrics, so post-quench evolution is not quasi-unitary in the initial physical space.
Choosing the preparation/measurement space as reference, we derive from first principles a normalized mixed-state Loschmidt amplitude using metric-compatible purifications.
The resulting biorthogonal trace representation contains a normalization factor given by the norm of the evolved purification in the initial metric; this factor is essential for comparing states from different physical Hilbert spaces and directly encodes the metric mismatch.
We prove that this normalization function is strictly positive on the real-time axis, so real-time criticality arises solely from destructive interference in the dynamical overlap.
In complex time, zeros of the normalization factor generate branch singularities distinct from Fisher zeros.
For a general two-level system, we derive analytic expressions for the Fisher and normalization zeros and analyze exceptional-point scaling together with the temperature- and quasi-Hermiticity-dependent DQPT critical times in two solvable quench protocols.
In contrast to metric-stationary quenches, where non-Hermitian effects may be absorbed into an equivalent Hermitian description, the metric mismatch here produces genuine non-Hermitian dynamics: a time-dependent normalization factor and complex-time branch structure that cannot be removed by any global similarity transformation.
\end{abstract}

\maketitle

\section{Introduction}

Dynamical quantum phase transitions (DQPTs) extend the connection between partition-function zeros and free-energy nonanalyticities to real-time evolution.
The Loschmidt amplitude acts as a dynamical partition function in complex time; in the thermodynamic limit, its Fisher zeros may condense into lines or surfaces, and nonanalyticities appear in the rate function when these structures cross the real-time axis~\cite{Fisher1967,Yang1952,Lee1952,Heyl13,Heyl_2018}.
This framework has been applied to interacting, topological, long-range, Floquet, and slow-quench systems, and has been probed in quantum simulators, ultracold atoms, and trapped ions~\cite{Karrasch2013,Kriel2014,Andraschko2014,Heyl2015,Sharma2016,PhysRevLett.120.130601,Jurcevic2017,Flaschner2018,Guo2019,Yang2019}.
DQPTs are also linked to geometric phases, dynamical topological order parameters, and momentum-space vortices~\cite{Budich2016,Vajna2015,Caio2015,Hu2016,Huang2016,Lang2018b}.

At finite temperature, interferometric Loschmidt amplitudes, mixed-state fidelities, and purification-based formulations have shown that thermal fluctuations can shift Fisher zeros and reshape phase-vortex structures without necessarily destroying criticality~\cite{Abeling2016,Heyl2017Mixed,Bhattacharya2017,Mera2018,Sedlmayr2018,Hou2020UbiquityZerosLoschmidt,Hou2022MetamorphicDynamicalQuantum,Tang2024UhlmannQuenchGeometric,Tang2025GeometryEffectDynamical}.
In non-Hermitian systems, nonorthogonal eigenstates and nonunitary evolution further require specifying dual states, inner-product structure, and normalization when defining the Loschmidt amplitude~\cite{Ashida2020,Brody2014}.
Early studies focused on zero-temperature pure-state models~\cite{Zhou2018,Zhou2021b,Naji2022b,Hamazaki2021,Mondal2022}; more recent works introduced self-normal, biorthogonal, and associated-state constructions to analyze zero conditions and topological responses~\cite{Jing2024,Zhang2025SelfNormal,gu2026biorthogonaldynamicalquantumphase}.
At finite temperature, thermal fluctuations substantially affect Fisher-zero distributions and winding numbers~\cite{Mondal2023}, and a unified framework now covers pure and mixed states within biorthogonal and self-norm formulations~\cite{92w1-71rt}.

Quasi-Hermitian quantum mechanics endows non-Hermitian systems with a consistent probabilistic interpretation through a positive-definite metric under which the Hamiltonian is self-adjoint and evolution is unitary~\cite{Mostafazadeh2002I,doi:10.1142/S0219887810004816,Bender1998}.
The metric is not fixed by the Hamiltonian alone but by the observables of state preparation and measurement, up to an irrelevant overall factor for a complete irreducible set~\cite{Scholtz1992QuasiHermitian,MostafazadehOzcelik2006,Geyer2007MetricObservables}.
As parameters vary, the compatible metric may change, altering physical overlaps and purification geometry~\cite{Wang2024InterferometricGeometricPhases,Hou2026QuasiHermitianUhlmann}.
For a sudden quench, post-quench evolution generally does not preserve the initial inner product, so the Loschmidt amplitude must be defined in a reference space fixed by preparation and readout.

We choose the initial preparation space as reference, where both the initial state and the return measurement are defined.
From metric-compatible purifications, we derive a first-principles expression for the mixed-state Loschmidt amplitude: the overlap between initial and evolved purifications, normalized by the evolved purification norm in the initial metric.
This gives a biorthogonal trace form whose normalization factor directly measures the deviation from quasi-unitarity and is indispensable for comparing states from different physical Hilbert spaces.
The positivity of the reference metric ensures this factor is strictly positive for real times, so real-time zeros arise only from destructive interference in the dynamical overlap.
In complex time, zeros of the normalization factor generate branch singularities distinct from Fisher zeros.

We apply the framework to two complementary two-level quenches: a gap-closing protocol approaching a second-order exceptional point and an isospectral protocol with fixed level spacing.

In our previous metric-stationary study with $\eta_i=\eta_f$, the Loschmidt amplitude reduced to that of an associated Hermitian system, making non-Hermitian effects appear removable~\cite{Tang2026ThermalSuppression}.
Here, metric mismatch breaks this equivalence: post-quench evolution is not quasi-unitary in the initial space, giving a time-dependent normalization and complex-time branch singularities without Hermitian counterpart, irremovable by global similarity transformations.

Together, the two protocols distinguish gap-closing scaling from metric-induced effects at fixed spectrum.

The remainder of this paper is organized as follows.
Section~\ref{sec:quench-protocol} introduces quasi-Hermitian mixed-state quenches, including the metric mismatch and the choice of the initial reference space.
Section~\ref{sec:purification-loschmidt} derives the normalized mixed-state Loschmidt amplitude from purification geometry.

Section~\ref{sec:complex-time} develops the complex-time theory of Fisher zeros, normalization zeros, and exceptional-point scaling.
Section~\ref{sec:two-level-applications} examines two complementary two-level quench protocols: a finite-temperature quench toward an exceptional point and an isospectral quench at fixed level spacing.

Section~\ref{sec:conclusion} summarizes our results.
Technical details are provided in the appendices.

\section{Quasi-Hermitian Mixed-State Quenches}
\label{sec:quench-protocol}

We begin by reviewing the essential structure of quasi-Hermitian mixed states and the metric mismatch generated by a sudden quench, which fixes the reference physical space used throughout this work.

\subsection{Basic Structure of Quasi-Hermitian Mixed States}
\label{subsec:quasi-Hermitian-mixed-states}

A finite-dimensional diagonalizable Hamiltonian $H$ is quasi-Hermitian if there exists a positive-definite Hermitian metric $\eta$ such that $H^\dagger=\eta H\eta^{-1}$.
The metric defines the physical inner product $(\psi,\phi)_\eta=\langle\psi|\eta|\phi\rangle$ and the adjoint operation $A^\ddagger=\eta^{-1}A^\dagger\eta$. Then $H^\ddagger=H$, so $H$ is self-adjoint in the physical Hilbert space $\mathcal H_\eta$ and has a real spectrum.
Its propagator $U(t)=\me^{-\mi Ht}$ satisfies $U^\ddagger(t)U(t)=I$,
so $U(t)$ is quasi-unitary and preserves the physical norm. The standard Dirac norm is generally not conserved, but probability normalization in $\mathcal H_\eta$ remains intact throughout the evolution~\cite{Mostafazadeh2002I,doi:10.1142/S0219887810004816}.
The Hamiltonian alone does not determine a unique compatible positive-definite metric; the latter is fixed by the complete set of observables associated with preparation and measurement.

For a non-Hermitian $H$, the spectral decomposition uses right eigenstates $|\psi_n\rangle$ and left eigenstates $|\phi_n\rangle=\eta|\psi_n\rangle$, which satisfy biorthogonality and completeness. Then $H=\sum_n E_n|\psi_n\rangle\langle\phi_n|$.
The physical trace of an operator $A$ is $\operatorname{Tr}_{\eta}(A)
=
\sum_n\langle\phi_n|A|\psi_n\rangle=\sum_n\langle\psi_n|\eta A|\psi_n\rangle$,
which in finite dimension equals the ordinary trace $\operatorname{Tr}(A)$. The metric does not alter the trace but fixes the inner product, duality, and adjoint structure of the physical Hilbert space.

A quasi-Hermitian mixed state is described by a density operator $\rho$ that is self-adjoint in the physical inner product, positive semidefinite, and normalized: $\rho^\ddagger=\rho$, $(\psi,\rho\psi)_\eta\geq0$, and $\operatorname{Tr}_{\eta}(\rho)=1$. These conditions generalize the standard density-matrix requirements to the quasi-Hermitian setting~\cite{Hou2026QuasiHermitianUhlmann}.
At temperature $T$, the thermal state is
\begin{align}
\rho_\beta
=
\frac{\me^{-\beta H}}{Z}
=
\sum_n p_n|\psi_n\rangle\langle\phi_n|,
\label{Gibbs-state}
\end{align}
with $\beta=(k_{\mathrm B}T)^{-1}$, $Z=\sum_n \me^{-\beta E_n}$, and $p_n=\me^{-\beta E_n}/Z$. As $T$ decreases, low-energy states acquire larger weights; at zero temperature, $\rho_\beta$ approaches the ground-state projector, or the normalized projector onto the ground-state subspace if degenerate. As $T$ increases, occupations become uniform, and in the infinite-temperature limit a finite-dimensional system approaches the maximally mixed state $I/d$, where $d$ is the Hilbert-space dimension.
Moreover, this thermal state admits a purification~\cite{Hou2026QuasiHermitianUhlmann}: $\rho_\beta = W_\beta W_\beta^{\ddagger}$, where
\begin{equation}
W_\beta = \sum_n \sqrt{p_n}\,|\psi_n\rangle\langle\phi_n|\,\mathcal{U},
\label{thermal-purification-general}
\end{equation}
and $\mathcal{U}$ is an arbitrary quasi-unitary operator satisfying $\mathcal{U}^{\ddagger}\mathcal{U}=I$, encoding the gauge freedom of the purification.

\subsection{Metric-Changing Quenches in Quasi-Hermitian Systems}
\label{subsec:metric-mismatch}

Consider a sudden quench from $H_{\mathrm i}$ to $H_{\mathrm f}$, with both Hamiltonians quasi-Hermitian: $H_\alpha^\dagger=\eta_\alpha H_\alpha\eta_\alpha^{-1}$, $\alpha=\mathrm i,\mathrm f$.
Before the quench, the system is prepared in the thermal state $\rho_{\mathrm i}=\me^{-\beta H_{\mathrm i}}/Z_{\mathrm i}$, whose physical inner product, self-adjointness, positivity, and normalization are defined by the initial metric $\eta_{\mathrm i}$. After the quench, the evolution is generated by $H_{\mathrm f}$; the corresponding time evolution $U_{\mathrm f}(t)=\me^{-\mi H_{\mathrm f}t}$ is quasi-unitary with respect to $\eta_{\mathrm f}$, i.e., $U_{\mathrm f}^{\ddagger_{\mathrm f}}(t)U_{\mathrm f}(t)=I$, with $A^{\ddagger_\alpha}=\eta_\alpha^{-1}A^\dagger\eta_\alpha$.

When $\eta_{\mathrm i}=\eta_{\mathrm f}$, the initial state and the post-quench evolution share the same physical Hilbert space, and the conventional mixed-state quench dynamics applies directly. In our previous work, we considered this metric-stationary regime and found that, under that specific condition, non-Hermitian effects may be absorbed into a global similarity transformation, so that the Loschmidt amplitude reduces to that of an associated Hermitian system with renormalized energies~\cite{Tang2026ThermalSuppression}. No essential non-Hermitian signature was observed in that setting. This suggests that a fixed metric may suppress distinctive non-Hermitian effects, and it motivates us to go beyond that restriction by considering quenches where the metric also changes, in which non-Hermiticity may become dynamically significant.

When $\eta_{\mathrm i}\neq\eta_{\mathrm f}$, however, the quench changes both the Hamiltonian and the physical inner product. The self-adjointness and positivity of $\rho_{\mathrm i}$ then hold only with respect to $\eta_{\mathrm i}$; in general, $\rho_{\mathrm i}^{\ddagger_{\mathrm f}}\neq\rho_{\mathrm i}$, and positivity with respect to $\eta_{\mathrm f}$ is not guaranteed. At the same time, $U_{\mathrm f}(t)$ preserves the $\eta_{\mathrm f}$ norm rather than the $\eta_{\mathrm i}$ norm.

The metric mismatch therefore places the initial state and the post-quench evolution in different physical Hilbert spaces. To define their overlap and the corresponding mixed-state Loschmidt amplitude, one must specify an identification between these spaces and represent both the initial state and the evolution in a common reference physical space.

\subsection{Reference Space and Non-Quasi-Unitary Evolution}
\label{subsec:reference-space-evolution}

Since the metric mismatch places the initial and post-quench states in different physical Hilbert spaces, a common reference space must be fixed before their overlap can be defined. In quasi-Hermitian quantum mechanics, this reference metric is not determined by the Hamiltonian alone: a complete physical model must also specify the observables $O_\mu$ associated with state preparation and measurement, all of which must be self-adjoint with respect to the same metric $\eta$, i.e., $O_\mu^\dagger\eta=\eta O_\mu$~\cite{Scholtz1992QuasiHermitian,MostafazadehOzcelik2006,Geyer2007MetricObservables}. If another positive-definite metric $\widetilde\eta$ is compatible with the same observable set, then $X=\eta^{-1}\widetilde\eta$ commutes with every $O_\mu$. When the observable set is complete and irreducible, its commutant contains only multiples of the identity, so $\widetilde\eta=c\eta$ with $c>0$, an overall factor that drops out of normalized physical quantities. The metric is therefore fixed by the full preparation and measurement protocol rather than by the Hamiltonian alone; it is part of the physical data, not a free gauge choice. This is precisely the situation in the examples below, where the additional observable $\sigma_y$ is used to select a unique metric $\eta(\lambda)=a(\lambda)I-b(\lambda)\sigma_y$ (with $\det\eta=1$) for the quasi-Hermitian Hamiltonian $H(\lambda)$.

In the quench considered here, this principle fixes $\eta_{\mathrm i}$ and $\eta_{\mathrm f}$ separately.
Although the pre- and post-quench states can be mapped to any fixed reference space, we choose the initial physical space $\mathcal H_{\eta_{\mathrm i}}$ because both the initial-state preparation and the return measurement are defined by the initial inner product.
This choice identifies the two Hilbert spaces without introducing an additional map; the final metric $\eta_{\mathrm f}$ remains relevant only for the self-adjointness of $H_{\mathrm f}$ in its own physical space.

However, in $\mathcal H_{\eta_{\mathrm i}}$ the adjoint of the final Hamiltonian is $H_{\mathrm f}^{\ddagger_{\mathrm i}}=\eta_{\mathrm i}^{-1}H_{\mathrm f}^\dagger\eta_{\mathrm i}$.
When $\eta_{\mathrm f}\neq\eta_{\mathrm i}$, one generally has $H_{\mathrm f}^{\ddagger_{\mathrm i}}\neq H_{\mathrm f}$, so $U_{\mathrm f}(t)=\me^{-\mi H_{\mathrm f}t}$ is not quasi-unitary with respect to $\eta_{\mathrm i}$.
The density operator evolves as
\begin{align}
\rho(t)
=
U_{\mathrm f}(t)\rho_{\mathrm i}
U_{\mathrm f}^{\ddagger_{\mathrm i}}(t)=
\me^{-\mi H_{\mathrm f}t}\rho_{\mathrm i}
\me^{\mi H_{\mathrm f}^{\ddagger_{\mathrm i}}t},
\label{density-matrix-evolution}
\end{align}
preserving self-adjointness and positivity with respect to $\eta_{\mathrm i}$, but not its physical trace:
\begin{align}
\operatorname{Tr}_{\eta_{\mathrm i}}\rho(t)
=
\operatorname{Tr}_{\eta_{\mathrm i}}
\left[
\rho_{\mathrm i}
U_{\mathrm f}^{\ddagger_{\mathrm i}}(t)
U_{\mathrm f}(t)
\right]
\neq 1 .
\label{evolved-density-weight}
\end{align}
This trace variation reflects the norm change generated by the post-quench dynamics relative to the initial measurement inner product; it does not imply probability loss in $\mathcal H_{\eta_{\mathrm f}}$.
The evolved state is amplified or attenuated when measured with the initial metric because the post-quench dynamics is quasi-unitary only with respect to $\eta_{\mathrm f}$, not $\eta_{\mathrm i}$.
%This non-quasi-unitarity necessitates a normalized definition of the mixed-state Loschmidt amplitude, which we develop next.

\section{Purification Construction of the Loschmidt Amplitude}
\label{sec:purification-loschmidt}

As shown in the previous section, the post-quench evolution is generally not quasi-unitary in the initial reference space. This necessitates a normalized definition of the mixed-state Loschmidt amplitude, which we now construct using the Hilbert--Schmidt inner product induced by the initial metric.

\subsection{Normalized Loschmidt Amplitude}
\label{subsec:normalized-purification-overlap}

In the initial physical space $\mathcal H_{\eta_{\mathrm i}}$, the thermal state admits a purification $\rho_{\mathrm i}=W(0)W^{\ddagger_{\mathrm i}}(0)$, where $W(0)$ is an amplitude in the enlarged Hilbert space. The mixed-state Loschmidt amplitude is then obtained as the overlap between the initial purification and its time-evolved counterpart, measured by the Hilbert--Schmidt inner product induced by $\eta_{\mathrm i}$~\cite{Uhlmann1986,Hou2026QuasiHermitianUhlmann,Tang2026ThermalSuppression}:
\begin{align}
(W_1,W_2)_{\mathrm{HS}}^{\eta_{\mathrm i}}
=
\operatorname{Tr}_{\eta_{\mathrm i}}
\left(
W_1^{\ddagger_{\mathrm i}}W_2
\right).
\label{eta-HS-inner-product}
\end{align}
The initial purification is normalized: $(W(0),W(0))_{\mathrm{HS}}^{\eta_{\mathrm i}}
=
\operatorname{Tr}_{\eta_{\mathrm i}}\rho_{\mathrm i}
=
1 $.
After the quench, the purification amplitude evolves as
\begin{align}
W(t)
=
U_{\mathrm f}(t)W(0)
=
\me^{-\mi H_{\mathrm f}t}W(0).
\label{purification-evolution}
\end{align}
The resulting density operator $\rho(t)=W(t)W^{\ddagger_{\mathrm i}}(t)$ agrees with the reference-space evolution in Eq.~\eqref{density-matrix-evolution}. Because $U_{\mathrm f}(t)$ is not quasi-unitary with respect to $\eta_{\mathrm i}$, the Hilbert--Schmidt norm of $W(t)$ is not conserved. We therefore define the mixed-state Loschmidt amplitude as the normalized overlap
\begin{align}
\mathcal G_\rho(t)
=
\frac{
(W(0),W(t))_{\mathrm{HS}}^{\eta_{\mathrm i}}
}{
\sqrt{
(W(0),W(0))_{\mathrm{HS}}^{\eta_{\mathrm i}}
(W(t),W(t))_{\mathrm{HS}}^{\eta_{\mathrm i}}
}
}.
\label{normalized-overlap-full}
\end{align}
Using the purification relation and cyclicity of the physical trace, we obtain
\begin{align}
(W(0),W(t))_{\mathrm{HS}}^{\eta_{\mathrm i}}
&=
\operatorname{Tr}_{\eta_{\mathrm i}}
\left(
\rho_{\mathrm i}\me^{-\mi H_{\mathrm f}t}
\right),
\nonumber\\
(W(t),W(t))_{\mathrm{HS}}^{\eta_{\mathrm i}}
&=
\operatorname{Tr}_{\eta_{\mathrm i}}
\left[
\rho_{\mathrm i}
\me^{\mi H_{\mathrm f}^{\ddagger_{\mathrm i}}t}
\me^{-\mi H_{\mathrm f}t}
\right].
\label{purification-overlaps}
\end{align}
With the initial purification normalized to unity, Eq.~\eqref{normalized-overlap-full} becomes
\begin{align}
\mathcal G_\rho(t)
=
\frac{
\operatorname{Tr}_{\eta_{\mathrm i}}
\left(
\rho_{\mathrm i}\me^{-\mi H_{\mathrm f}t}
\right)
}{
\sqrt{
\operatorname{Tr}_{\eta_{\mathrm i}}
\left[
\rho_{\mathrm i}
\me^{\mi H_{\mathrm f}^{\ddagger_{\mathrm i}}t}
\me^{-\mi H_{\mathrm f}t}
\right]
}
}.
\label{metric-compatible-Loschmidt}
\end{align}
This definition is independent of the purification gauge: under any quasi-unitary right transformation $W(0)\to W(0)V$ and $W(t)\to W(t)V$ with $VV^{\ddagger_{\mathrm i}}=I$, both numerator and denominator remain invariant. Thus $\mathcal G_\rho(t)$ depends only on the initial mixed state, the post-quench evolution, and the reference physical inner product.

\subsection{Overlap and Normalization}
\label{subsec:overlap-normalization}

To separate the roles of the numerator and denominator in Eq.~\eqref{metric-compatible-Loschmidt}, we write $\mathcal G_\rho(t)=N(t)/\sqrt{D(t)}$, where
\begin{align}
N(t)
&=
\operatorname{Tr}_{\eta_{\mathrm i}}
\left(
\rho_{\mathrm i}\me^{-\mi H_{\mathrm f}t}
\right),
\notag\\
D(t)
&=
\operatorname{Tr}_{\eta_{\mathrm i}}
\left[
\rho_{\mathrm i}
\me^{\mi H_{\mathrm f}^{\ddagger_{\mathrm i}}t}
\me^{-\mi H_{\mathrm f}t}
\right]
=
\operatorname{Tr}_{\eta_{\mathrm i}}\rho(t)
=
\|W(t)\|_{\mathrm{HS},\eta_{\mathrm i}}^{2}.
\label{real-time-denominator}
\end{align}
Here $N(t)$ is the unnormalized dynamical overlap, and $D(t)$ measures the norm variation of the evolved purification induced by the post-quench evolution relative to the initial metric. At $t=0$, $N(0)=D(0)=1$, so $\mathcal G_\rho(0)=1$.

Since $\eta_{\mathrm i}$ is positive definite, the Hilbert--Schmidt inner product obeys the Cauchy--Schwarz inequality. Applied to $W(0)$ and $W(t)$, this gives
\begin{align}
|N(t)|^2
&=
\left|
(W(0),W(t))_{\mathrm{HS}}^{\eta_{\mathrm i}}
\right|^2
\nonumber\\
&\le
(W(0),W(0))_{\mathrm{HS}}^{\eta_{\mathrm i}}
(W(t),W(t))_{\mathrm{HS}}^{\eta_{\mathrm i}}
=
D(t).
\label{HS-Cauchy-Schwarz}
\end{align}
For real time, $U_{\mathrm f}(t)$ is invertible and $D(t)>0$, as shown in Appendix~\ref{app:positive-D}. The normalization in Eq.~\eqref{metric-compatible-Loschmidt} is therefore well defined, and we obtain
\begin{align}
|\mathcal G_\rho(t)|^2
=
\frac{|N(t)|^2}{D(t)}
\le 1.
\label{Loschmidt-bound}
\end{align}

If $H_{\mathrm f}^{\ddagger_{\mathrm i}}=H_{\mathrm f}$, the post-quench evolution is quasi-unitary with respect to the initial metric, so $D(t)=1$ and Eq.~\eqref{metric-compatible-Loschmidt} reduces to the conventional mixed-state Loschmidt amplitude~\cite{Heyl2017Mixed,Bhattacharya2017,Mera2018,Sedlmayr2018}.

\section{Complex-Time Zeros and Branch Structure}
\label{sec:complex-time}

We now continue $N(t)$ and $D(t)$ to complex time $z\in\mathbb C$. The normalized amplitude becomes
\begin{align}
\mathcal G_\rho(z)
=
\frac{N(z)}{\sqrt{D(z)}},
\label{complex-loschmidt-decomposition}
\end{align}
where zeros of $N(z)$ are Fisher zeros, and zeros of $D(z)$ determine the branch structure of $\mathcal G_\rho(z)$.

\subsection{Fisher Zeros}
\label{subsec:fisher-zeros}

For a finite-dimensional system, $N(z)$ is an entire function. Since $N(0)=1$, it is not identically zero, and all its zeros are isolated Fisher zeros. Inserting the completeness relation of the final biorthogonal eigenbasis into the spectral decomposition of the initial thermal state gives
\begin{align}
N(z)
=
\sum_{n,m}
p_n \me^{-\mi E_m^{(\mathrm f)}z}
\langle\phi_n^{(\mathrm i)}|\psi_m^{(\mathrm f)}\rangle
\langle\phi_m^{(\mathrm f)}|\psi_n^{(\mathrm i)}\rangle .
\label{N-double-spectral-form}
\end{align}
Thus, Fisher zeros are determined by the initial thermal weights, the overlaps between pre- and post-quench eigenbases, and the final spectrum. Temperature changes $p_n$, redistributing contributions from different initial eigenstates and shifting the Fisher zeros, potentially driving them across the real-time axis. A real-time critical time $t_c$ satisfies
\begin{align}
\mathcal G_\rho(t_c)=0
\quad\Longleftrightarrow\quad
N(t_c)=0 ,
\label{real-time-zero-condition}
\end{align}
since the normalization factor is strictly positive on the real-time axis.

\subsection{Normalization Zeros and Branch Points}
\label{subsec:normalization-zeros}

Although $D(t)>0$ on the real-time axis, its complex-time continuation $D(z)$ is generally complex. For $z\in\mathbb C$,
$(\me^{-\mi H_{\mathrm f}z})^{\ddagger_{\mathrm i}}
=
\me^{\mi H_{\mathrm f}^{\ddagger_{\mathrm i}}z^*}$,
whereas the corresponding factor in $D(z)$ is $\me^{\mi H_{\mathrm f}^{\ddagger_{\mathrm i}}z}$; the two coincide only for $z\in\mathbb R$. Thus $D(z)$ is no longer a squared Hilbert--Schmidt norm and is not constrained by positivity. Different spectral contributions can cancel, generating complex-time zeros distinct from Fisher zeros.

Let $z_d$ be a zero of $D(z)$ of order $\nu_d$. In its neighborhood,
\begin{align}
D(z)
=
(z-z_d)^{\nu_d}d_d(z),
\qquad
d_d(z_d)\neq0 .
\label{local-D-zero}
\end{align}
The local behavior of $\sqrt{D(z)}$ is governed by $(z-z_d)^{\nu_d/2}$: odd $\nu_d$ yields a branch point, while even $\nu_d$ allows a single-valued branch. If $N(z)$ has a zero of order $\mu_d$ at the same point, with $N(z)=(z-z_d)^{\mu_d}n_d(z)$ and $n_d(z_d)\neq0$, then
\begin{align}
\mathcal G_\rho(z)
\sim
(z-z_d)^{\mu_d-\nu_d/2}
\frac{n_d(z_d)}{\sqrt{d_d(z_d)}} .
\label{local-normalized-amplitude}
\end{align}
Thus the local behavior of $\mathcal G_\rho(z)$ is controlled by $\mu_d-\nu_d/2$; a numerator zero can modify the power but cannot remove an odd-order branch point. When $\nu_d$ is odd, $z_d$ always remains a branch point; when $\nu_d$ is even, the singularity may be a pole, a removable singularity, or a zero depending on the relative orders.

Because $D(t)>0$, zeros of $D(z)$ cannot lie on the real-time axis. If $H_{\mathrm f}^{\ddagger_{\mathrm i}}=H_{\mathrm f}$, then $D(z)=1$, and neither normalization zeros nor the associated branch structure occur. The branch points are therefore a direct signature of the metric mismatch: when $\eta_i=\eta_f$, $\mathcal{G}_\rho(z)$ reduces to the entire function $N(z)$, so the branch structure encodes the non-quasi-unitary character of the post-quench evolution in the reference space.

\subsection{Exceptional-Point Scaling}
\label{sec:general-two-level-ep}

For a two-level quasi-Hermitian system, the spectral decompositions of $N(z)$ and $D(z)$ yield closed forms, allowing the two classes of complex-time zeros and their exceptional-point scaling to be characterized without specifying a particular quench protocol.

\subsubsection{Fisher and Normalization Zeros}
\label{subsec:two-level-complex-zeros}

Let the eigenvalues of the diagonalizable final Hamiltonian $H_{\mathrm f}$ be $E_\pm^{(\mathrm f)}=\bar E_{\mathrm f}\pm\Delta_{\mathrm f}$, with gap $2\Delta_{\mathrm f}$, and let
$P_\pm^{(\mathrm f)}=|\psi_\pm^{(\mathrm f)}\rangle\langle\phi_\pm^{(\mathrm f)}|$
be the corresponding biorthogonal projectors. The spectral decomposition then gives
\begin{align}
N(z)
=
\me^{-\mi \bar E_{\mathrm f}z}
\left(
C_+\me^{-\mi \Delta_{\mathrm f}z}
+
C_-\me^{\mi \Delta_{\mathrm f}z}
\right),
\label{two-level-N}
\end{align}
where $C_\pm=\operatorname{Tr}_{\eta_{\mathrm i}}[\rho_{\mathrm i}P_\pm^{(\mathrm f)}]$ and $C_++C_-=1$. The coefficients $C_\pm$ encode the initial thermal weights and the biorthogonal overlaps between pre- and post-quench eigenstates; the mean energy contributes only a global phase and does not affect Fisher-zero locations.

For $C_+C_-\neq0$, solving $N(z)=0$ yields
\begin{align}
z_k^{\mathrm F}
=
\frac{\operatorname{Arg}(-C_+/C_-)+2\pi k}
{2\Delta_{\mathrm f}}
-
\frac{\mi}{2\Delta_{\mathrm f}}
\ln\left|
\frac{C_+}{C_-}
\right|.
\label{general-fisher-zeros}
\end{align}
Here $k\in\mathbb Z$ labels the Fisher zeros generated by phase periodicity; they are equally spaced along the real-time direction. Their real parts are fixed by the relative phase of the two spectral contributions, while their imaginary parts reflect the magnitude difference. A Fisher zero lies on the real-time axis when $|C_+|=|C_-|$; with $C_++C_-=1$, this is equivalent to $\operatorname{Re}C_+=1/2$. If $C_+=0$ or $C_-=0$, no finite complex-time Fisher zero exists.

The same spectral decomposition gives the normalization function
\begin{align}
D(z)
=
D_0
+
D_+\me^{2\mi\Delta_{\mathrm f}z}
+
D_-\me^{-2\mi\Delta_{\mathrm f}z},
\label{general-D}
\end{align}
where $D_0$, $D_+$, and $D_-$ are specified in Appendix~\ref{app:two-level-coefficients}, together with the degenerate cases. For $D_+D_-\neq0$, define
\begin{align}
y_\pm
=
\frac{-D_0\pm\sqrt{D_0^2-4D_+D_-}}
{2D_+},
\end{align}
so the normalization zeros are
\begin{align}
z_{m,\pm}^{\mathrm D}
=
\frac{\operatorname{Arg}y_\pm+2\pi m}
{2\Delta_{\mathrm f}}
-
\frac{\mi}{2\Delta_{\mathrm f}}
\ln|y_\pm|,
\label{general-D-zeros}
\end{align}
with $m\in\mathbb Z$. Since $D(t)>0$ for real times, all these zeros lie away from the real-time axis. When $D_0^2-4D_+D_-\neq0$, they are simple zeros of $D(z)$ and generate square-root branch points of $\sqrt{D(z)}$. If $H_{\mathrm f}^{\ddagger_{\mathrm i}}=H_{\mathrm f}$, then $D(z)=1$ and no normalization zeros arise.

\subsubsection{Scaling Near an Exceptional Point}
\label{subsec:exceptional-point-scaling}

The complex-time zeros acquire particular physical significance near spectral degeneracies, where the non-Hermitian nature of the dynamics is amplified. Suppose the final Hamiltonian $H_{\mathrm f}(\lambda)$ depends on a control parameter $\lambda$ and develops an exceptional degeneracy at $\lambda=\lambda_{\mathrm{EP}}$. In a two-level system this is a second-order exceptional point where both eigenvalues and eigenstates coalesce. Approaching the exceptional point from the unbroken side, the level splitting typically scales as
\begin{align}
2\Delta_{\mathrm f}
=
E_+^{(\mathrm f)}-E_-^{(\mathrm f)}
\propto
|\lambda-\lambda_{\mathrm{EP}}|^{1/2},
\label{ep-gap-closing}
\end{align}
so the post-quench gap closes with a square-root law~\cite{Heiss12,Bergholtz2021}. Equations~\eqref{general-fisher-zeros} and~\eqref{general-D-zeros} show that both classes of complex-time zeros contain an explicit factor $1/\Delta_{\mathrm f}$. If the dimensionless phase or logarithmic term in the corresponding zero formula approaches a finite nonzero constant as $\Delta_{\mathrm f}\to0$, then a fixed zero branch obeys
\begin{align}
|z_k^{\mathrm F}|,\ |z_{m,\pm}^{\mathrm D}|
\sim
\frac{1}{|\Delta_{\mathrm f}|}
\propto
|\lambda-\lambda_{\mathrm{EP}}|^{-1/2}.
\label{ep-time-scale}
\end{align}
Thus, as the post-quench gap closes, the complex-time zeros on a fixed branch move progressively away from the origin. This divergence reflects the increasing sensitivity of the interference conditions near the exceptional point, where the biorthogonal spectral description breaks down.

Under the stated conditions, the two classes of zeros share the same leading gap scaling but have different origins: Fisher zeros arise from cancellation between the two spectral contributions to $N(z)$, whereas normalization zeros arise from zeros of the two-sided evolution function $D(z)$; their locations generally do not coincide. Equation~\eqref{ep-time-scale} is not unconditional. Near the exceptional point, the spectral projectors, the positive-definite metric, and the coefficients $C_\pm$, $D_0$, and $D_\pm$ may themselves acquire singular gap dependence, which can modify or even invalidate the leading $1/|\Delta_{\mathrm f}|$ scaling. At the exceptional point, $H_{\mathrm f}$ is no longer diagonalizable, and the spectral expansion in terms of biorthogonal eigenprojectors ceases to apply; the propagator must then be reconstructed from the Jordan canonical form.

\section{Two-Level Quench Protocols}
\label{sec:two-level-applications}

To illustrate the general theory, we consider two exactly solvable two-level quench protocols with complementary spectral properties. The first is a gap-closing quench toward a second-order exceptional point and is used to characterize the scaling of complex-time zeros. The second is isospectral and examines how temperature and quasi-Hermitian parameters modify the DQPT critical times while the level spacing remains fixed.

\subsection{Finite-Temperature Quench toward an Exceptional Point}
\label{subsec:explicit-ep-example}

Consider the one-parameter two-level Hamiltonian
\begin{align}
H(g)=\Omega\left(\sigma_x+\mi g\sigma_z\right),
\qquad 0\le g<1.
\label{explicit-ep-family}
\end{align}
It obeys $H^2(g)=\Omega^2(1-g^2)I$,
and therefore has eigenvalues $E_\pm(g)=\pm\Omega\sqrt{1-g^2}$.
The spectrum is real and nondegenerate for $0\le g<1$, while the two eigenvalues coalesce at $g=1$.

The physical inner product is fixed by including $\sigma_y$ in the observable set and requiring both $H(g)$ and $\sigma_y$ to be self-adjoint with respect to the same metric. Since $\sigma_y^\dagger=\sigma_y$, the condition $\sigma_y^\dagger\eta=\eta\sigma_y$
implies $[\eta,\sigma_y]=0$, so the metric may be written as
\begin{align}
\eta(g)=u(g)I+v(g)\sigma_y.
\end{align}
The quasi-Hermiticity condition $H^\dagger(g)\eta(g)=\eta(g)H(g)$
gives $v=gu$. Fixing the overall normalization by $\det\eta(g)=1$ then yields
\begin{align}
\eta(g)=\frac{I+g\sigma_y}{\sqrt{1-g^2}}.
\label{explicit-ep-metric}
\end{align}
Its eigenvalues, $\eta_\pm(g)=\frac{1\pm g}{\sqrt{1-g^2}}$,
are positive for $0\le g<1$, so $H(g)$ is quasi-Hermitian throughout this interval.

At $g=1$, $H^2(1)=0$ and $H(1)\neq0$,
while the zero-eigenvalue eigenspace is one dimensional,
\begin{align}
\ker H(1)
=
\operatorname{span}
\left\{
\begin{pmatrix}
1\\
-\mi
\end{pmatrix}
\right\}.
\end{align}
Thus the eigenvalue $E=0$ has algebraic multiplicity two but geometric multiplicity one. The Hamiltonian is therefore defective at $g=1$, identifying this point as a second-order exceptional point.

\begin{figure}[t]
\centering
\includegraphics[width=0.48\textwidth]{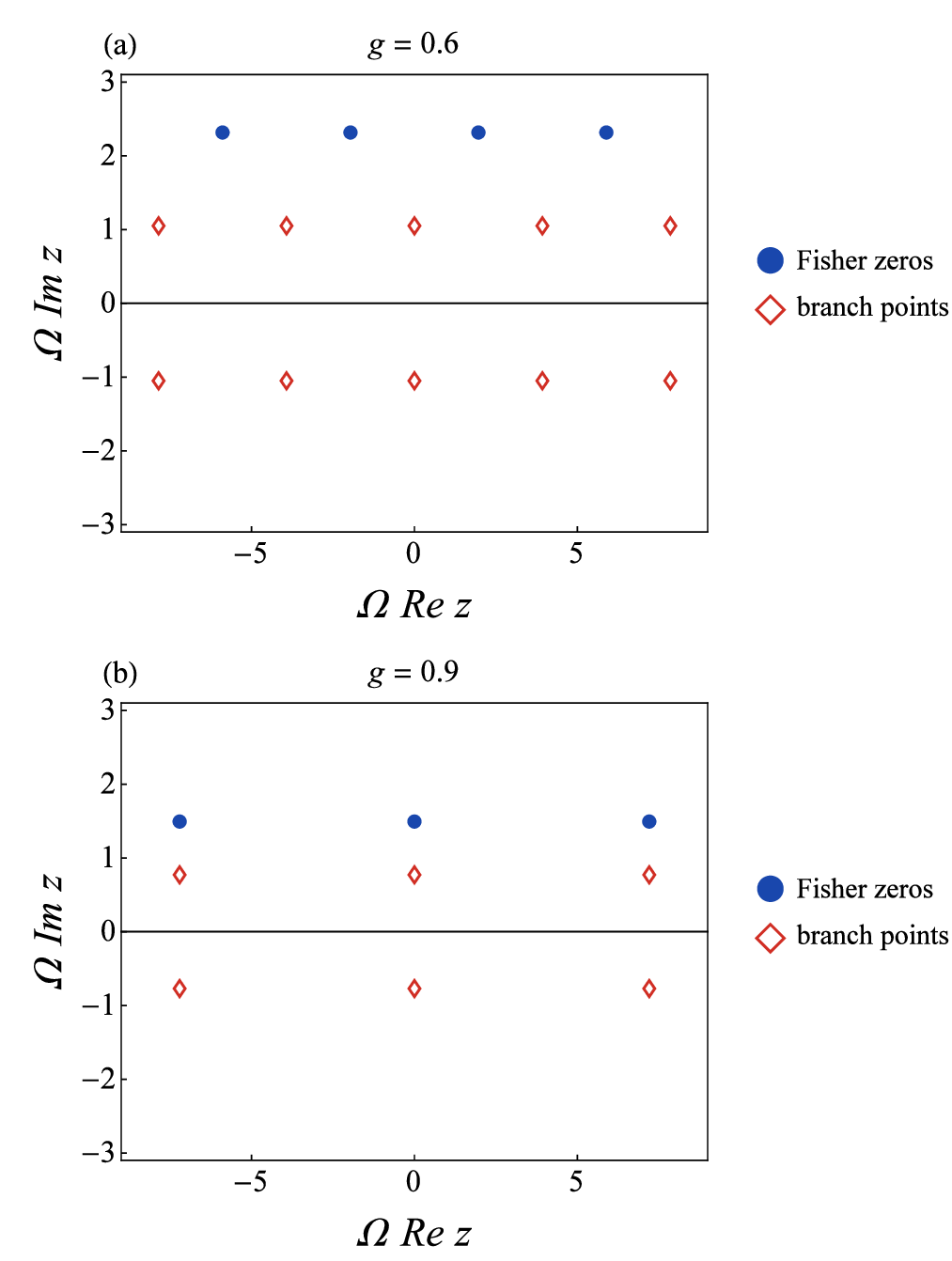}
\caption{Complex-time Fisher zeros and normalization branch points for the finite-temperature quench $g_i=0\to g_f=g$ at $\beta\Omega=1$. (a) $g=0.6$ and (b) $g=0.9$. Blue solid circles denote Fisher zeros, and red open diamonds denote normalization branch points.}
\label{fig:explicit-ep-complex-zeros}
\end{figure}

We consider a sudden quench from $g_i=0$ to $g_f=g$. The corresponding initial Hamiltonian and metric are $H_{\mathrm i}=\Omega\sigma_x$ and $\eta_{\mathrm i}=I$ respectively,
and the initial state is the Gibbs state
\begin{align}
\rho_{\mathrm i}
=
\frac{\me^{-\beta H_{\mathrm i}}}{Z_{\mathrm i}}
=
\frac12\left(I-\chi\sigma_x\right),
\qquad
\chi=\tanh(\beta\Omega).
\label{explicit-ep-thermal-state}
\end{align}
For the post-quench Hamiltonian $H_{\mathrm f}=H(g)$, define $\delta=\sqrt{1-g^2}$ and $\Delta_{\mathrm f}=\Omega\delta$.
Using $H_{\mathrm f}^2=\Delta_{\mathrm f}^2I$, the complex-time propagator takes the form
\begin{align}
\me^{-\mi H_{\mathrm f}z}
=
\cos(\Delta_{\mathrm f}z)I
-\mi\frac{\sin(\Delta_{\mathrm f}z)}{\delta}
\left(\sigma_x+\mi g\sigma_z\right).
\label{explicit-ep-propagator}
\end{align}
The numerator and normalization function are then
\begin{align}
N(z)
&=
\cos(\Delta_{\mathrm f}z)
+\mi\frac{\chi}{\delta}\sin(\Delta_{\mathrm f}z),
\notag\\
D(z)
&=
1+\frac{2g^2}{1-g^2}\sin^2(\Delta_{\mathrm f}z).
\label{explicit-ep-D}
\end{align}
In this model, the temperature dependence enters $N(z)$ through the thermal polarization $\chi$, whereas $D(z)$ is independent of temperature.

For $\chi\neq\delta$, define
\begin{align}
r_\beta(g)=\frac{\chi-\delta}{\chi+\delta}.
\label{explicit-ep-rbeta}
\end{align}
The Fisher-zero condition $N(z)=0$ becomes $\me^{2\mi\Delta_{\mathrm f}z}=r_\beta(g)$,
and hence
\begin{align}
z_k^{\mathrm F}
=
\frac{\operatorname{Arg}r_\beta+2\pi k}{2\Delta_{\mathrm f}}
-\frac{\mi}{2\Delta_{\mathrm f}}\ln|r_\beta|,
\qquad
k\in\mathbb Z.
\label{explicit-ep-fisher-zeros}
\end{align}
At fixed temperature, the sign of $r_\beta$ changes at
\begin{align}
g_{\mathrm F}^{*}(\beta)
=
\sqrt{1-\chi^2}
=
\operatorname{sech}(\beta\Omega).
\label{explicit-ep-fisher-crossover}
\end{align}
For $g<g_{\mathrm F}^{*}(\beta)$, the Fisher zeros lie on the half-integer real-part lattice, $\operatorname{Re}z_k^{\mathrm F}
=
\frac{(k+1/2)\pi}{\Delta_{\mathrm f}}$,
whereas for $g>g_{\mathrm F}^{*}(\beta)$ they lie on the integer lattice, $\operatorname{Re}z_k^{\mathrm F}
=
\frac{k\pi}{\Delta_{\mathrm f}}$.
As $g\to g_{\mathrm F}^{*}(\beta)^{\pm}$, $|r_\beta|\to0$ and $\operatorname{Im}z_k^{\mathrm F}\to+\infty$.
At $g=g_{\mathrm F}^{*}(\beta)$ one spectral contribution to $N(z)$ vanishes, leaving a single exponential term and therefore no finite Fisher zero. The two real-part lattices are thus connected through complex infinity rather than by a continuous finite displacement.

The normalization zeros follow from Eq.~\eqref{explicit-ep-D}:
\begin{align}
z_{m,\pm}^{\mathrm D}
=
\frac{
m\pi
\pm
\mi\,\operatorname{arsinh}
\!\left[\delta/(\sqrt2 g)\right]
}{
\Delta_{\mathrm f}
},
\qquad
m\in\mathbb Z.
\label{explicit-ep-D-zeros}
\end{align}
For $0<g<1$, these are simple zeros of $D(z)$ and hence square-root branch points of $\sqrt{D(z)}$. They lie off the real-time axis and therefore do not generate real-time DQPTs; they control only the complex-time branch structure of $\mathcal{G}_\rho(z)$. In the metric-stationary limit $\eta_{\mathrm i}=\eta_{\mathrm f}$, one has $D(z)=1$ identically, so these zeros and the associated branch points are absent. Their appearance here is thus a direct fingerprint of the metric mismatch. Moreover,
\begin{align}
N(z_{m,\pm}^{\mathrm D})
=
\frac{(-1)^m}{\sqrt2g}
\left[
\sqrt{1+g^2}\mp\chi
\right]
\neq0,
\end{align}
for $0\le\chi\le1$. Consequently,
\begin{align}
\mathcal G_\rho(z)
\sim
\left(z-z_{m,\pm}^{\mathrm D}\right)^{-1/2}
\end{align}
in the vicinity of each normalization zero.

Figure~\ref{fig:explicit-ep-complex-zeros} illustrates these two regimes at $\beta\Omega=1$, for which $g_{\mathrm F}^{*}(\beta)
=
\operatorname{sech}(1)
\simeq0.648$.
For $g=0.6<g_{\mathrm F}^{*}(\beta)$, the Fisher zeros lie on the half-integer real-part lattice, whereas for $g=0.9>g_{\mathrm F}^{*}(\beta)$ they lie on the integer lattice. The normalization zeros form complex-conjugate pairs about the real axis and are independent of $\chi$.

\begin{figure}[t]
\centering
\includegraphics[width=0.48\textwidth]{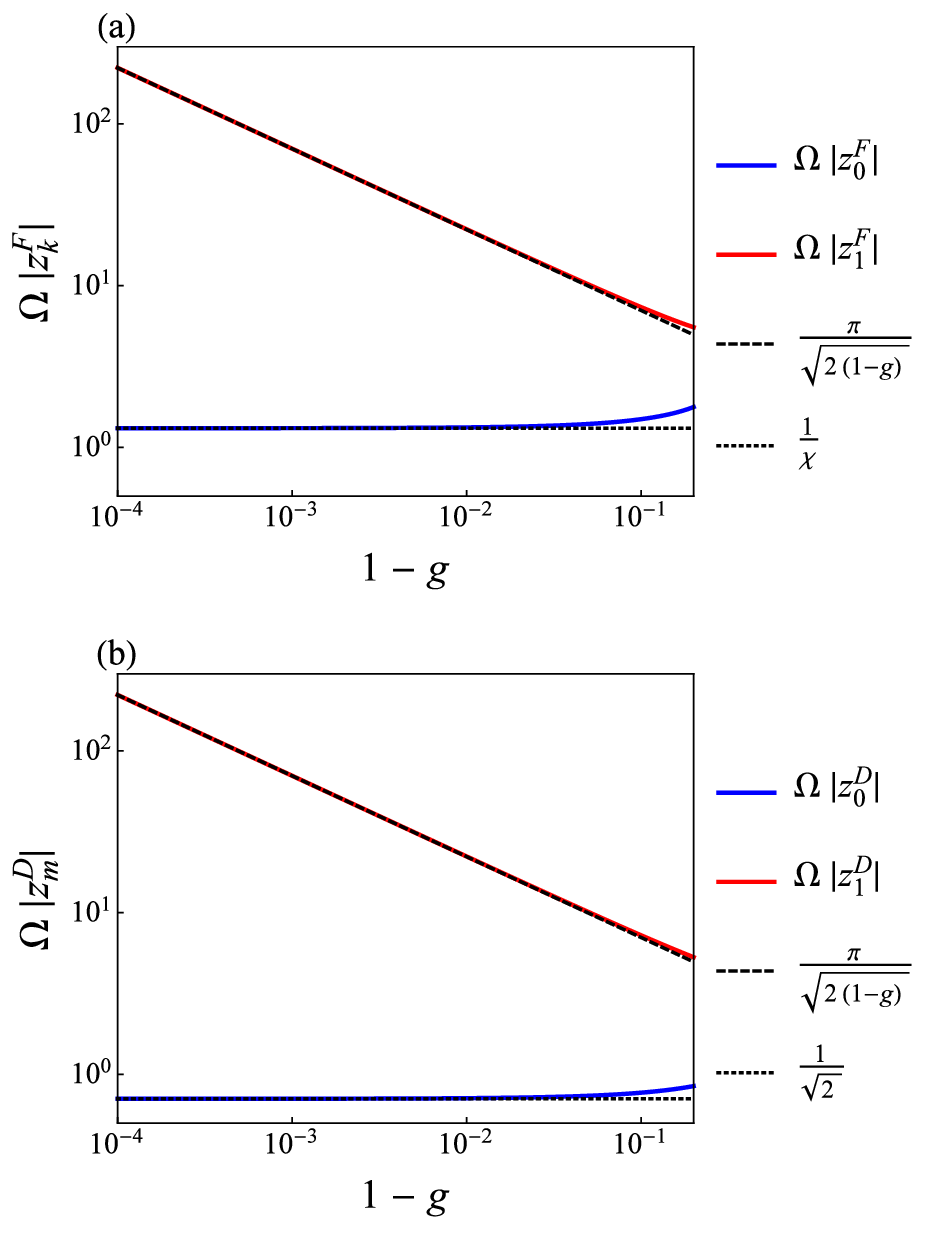}
\caption{Exceptional-point scaling of the complex-time zeros at $\beta\Omega=1$ as functions of $1-g$. (a) Fisher-zero branches $\Omega|z_0^{\mathrm F}|$ and $\Omega|z_1^{\mathrm F}|$. (b) Normalization-zero branches $\Omega|z_0^{\mathrm D}|$ and $\Omega|z_1^{\mathrm D}|$. Black dashed curves show $\pi/\sqrt{2(1-g)}$, while the black dotted curves show $1/\chi$ in (a) and $1/\sqrt2$ in (b).}
\label{fig:explicit-ep-scaling}
\end{figure}

The exceptional-point asymptotics follow directly from Eqs.~\eqref{explicit-ep-fisher-zeros} and~\eqref{explicit-ep-D-zeros}. As $g\to1^-$, $\delta
=
\sqrt{1-g^2}
\simeq
\sqrt{2(1-g)}$,
so that
\begin{align}
\frac{1}{\delta}
\propto
(1-g)^{-1/2}.
\end{align}
For any fixed finite temperature, $\chi>0$, and sufficiently close to the exceptional point one has $g>g_{\mathrm F}^{*}(\beta)$. Expanding Eq.~\eqref{explicit-ep-rbeta} gives $\ln r_\beta
=
-\frac{2\delta}{\chi}
+
O(\delta^3)$,
and therefore
\begin{align}
\Omega z_k^{\mathrm F}
=
\frac{k\pi}{\delta}
+
\frac{\mi}{\chi}
+
O(\delta^2).
\label{explicit-ep-fisher-asymptotic}
\end{align}
For $k\neq0$, $\Omega|z_k^{\mathrm F}|
\sim
\frac{|k|\pi}{\delta}
\propto
(1-g)^{-1/2}$,
whereas the central Fisher branch remains finite,
\begin{align}
\Omega z_0^{\mathrm F}
\longrightarrow
\frac{\mi}{\chi}.
\label{explicit-ep-fisher0-limit}
\end{align}

Similarly,
\begin{align}
\Omega z_{m,\pm}^{\mathrm D}
=
\frac{m\pi}{\delta}
\pm
\frac{\mi}{\sqrt2}
+
O(\delta^2),
\label{explicit-ep-D-asymptotic}
\end{align}
so that the $m\neq0$ normalization branches obey the same inverse-square-root scaling, while
\begin{align}
\Omega z_{0,\pm}^{\mathrm D}
\longrightarrow
\pm\frac{\mi}{\sqrt2}.
\label{explicit-ep-D0-limit}
\end{align}

Figure~\ref{fig:explicit-ep-scaling} compares the central and noncentral zero branches. In panel (a), $\Omega|z_1^{\mathrm F}|\propto(1-g)^{-1/2}$, whereas $\Omega|z_0^{\mathrm F}|\to1/\chi$. Panel (b) shows the analogous behavior for the normalization zeros: the $m=1$ branch has the same inverse-square-root scaling, while $\Omega|z_0^{\mathrm D}|\to1/\sqrt2$. Since $z_{m,+}^{\mathrm D}$ and $z_{m,-}^{\mathrm D}$ are complex conjugates, their moduli coincide and only one modulus branch is shown for each $m$. The contrast between finite and divergent branches confirms that the explicit factor $1/\Delta_{\mathrm f}$ yields inverse-gap divergence only when its dimensionless prefactor approaches a finite nonzero constant as $\Delta_{\mathrm f}\to0$.

This protocol thus exhibits four physical features. (i) Temperature acts only on the Fisher numerator, selecting the half-integer or integer real-part lattice of the Fisher zeros and driving the crossover at $g_{\mathrm F}^{*}(\beta)=\operatorname{sech}(\beta\Omega)$. (ii) Metric mismatch generates normalization zeros that are absent when $\eta_{\mathrm i}=\eta_{\mathrm f}$ and that depend only on $g$, not on temperature; being off the real-time axis, they generate square-root branch points of $\mathcal{G}_\rho(z)$ rather than real-time DQPTs. (iii) Gap closing drives the noncentral Fisher and normalization branches to diverge as $(1-g)^{-1/2}$, while the central branches remain finite, in agreement with the conditional scaling derived in Sec.~\ref{sec:general-two-level-ep}. (iv) Since the normalization zeros and the associated branch structure vanish identically in the metric-stationary limit, they cannot be removed by any global similarity transformation and constitute a genuine non-Hermitian effect induced by metric mismatch.

The isospectral protocol considered next focuses on the temperature- and quasi-Hermiticity-dependent DQPT critical times at fixed spectrum.

\subsection{Isospectral Quench Dynamics}
\label{sec:single-parameter-model}

The second protocol is an exactly solvable isospectral two-level model with fixed level spacing but parameter-dependent eigenstates and metric. It therefore isolates metric- and temperature-induced changes in the Loschmidt amplitude from spectral-gap effects.

\subsubsection{Model and Metric}
\label{subsec:model-metric-structure}

Consider the single-parameter two-level Hamiltonian
\begin{align}
H(\lambda)=\Omega\left[a(\lambda)\boldsymbol n_r(\lambda)+\mi b(\lambda)\boldsymbol n_\theta(\lambda)\right]\cdot\boldsymbol\sigma,
\label{single-parameter-H}
\end{align}
where $\Omega>0$, $\boldsymbol n_r=(\sin\lambda,0,\cos\lambda)$, and $\boldsymbol n_\theta=(\cos\lambda,0,-\sin\lambda)$ are mutually orthogonal unit vectors. We choose
\begin{align}
a(\lambda)=\sqrt{1+\gamma^2\sin^2\lambda},\qquad b(\lambda)=\gamma\sin\lambda.
\label{single-parameter-ab}
\end{align}
The parameter $\gamma\geq0$ controls the strength of the non-Hermitian term. Since $a^2-b^2=1$, the Hamiltonian obeys $H^2(\lambda)=\Omega^2I$ and $E_\pm(\lambda)=\pm\Omega$. Thus $\lambda$ changes only the eigenstates and the compatible metric, leaving the level spacing fixed; this isolates the effect of the metric on the critical times from any variation of the gap.

We take $\sigma_y$ as an additional observable and require both $\sigma_y$ and $H(\lambda)$ to be self-adjoint in the same physical inner product. Together with $\det\eta(\lambda)=1$, this fixes
\begin{align}
\eta(\lambda)=a(\lambda)I-b(\lambda)\sigma_y,
\label{single-parameter-metric}
\end{align}
as derived in Appendix~\ref{app:model-metric}.

With $\vartheta(\lambda)=\operatorname{arsinh}[\gamma\sin\lambda]$, the metric eigenvalues are $a\pm b=\me^{\pm\vartheta}>0$, so the metric is strictly positive definite for finite parameters. The parameter $\gamma$ thus controls both the non-Hermitian component of $H$ and the relative weights of the two $\sigma_y$ eigen-directions in the physical inner product.

We consider a sudden quench from $\lambda_i=0$ to $\lambda_f=\lambda$. At the initial point, $H_i=\Omega\sigma_z$, $\eta_i=I$, and $\rho_i=\frac12\left(I-\chi\sigma_z\right)$ with $\chi=\tanh(\beta\Omega)$. Here $\chi$ is the initial thermal polarization; $\chi\to1$ and $\chi\to0$ correspond to zero and infinite temperature, respectively. After the quench, the system evolves under $H_f=H(\lambda)$.

\subsubsection{Real-Time Critical Condition}
\label{subsec:real-time-fisher-zeros}

Since $D(t)>0$ for real time, real-time zeros of $\mathcal G_\rho(t)$ coincide with zeros of the Fisher numerator $N_\lambda(t)$. For the quench in Eq.~\eqref{single-parameter-H},
\begin{align}
N_\lambda(t)
&=
\cos(\Omega t)+\gamma\chi\sin^2\lambda\,\sin(\Omega t)
\nonumber\\
&\quad
+\mi\chi a(\lambda)\cos\lambda\,\sin(\Omega t).
\label{single-parameter-numerator}
\end{align}
At finite temperature, $\chi>0$, so $N_\lambda(t)=0$ requires $\cos\lambda_f=0$. Hence, finite-temperature DQPTs occur only for orthogonal quenches, $\lambda_f=\pi/2\pmod{\pi}$; for non-orthogonal quenches, no real-time Fisher zeros exist at finite temperature. For an orthogonal quench, the critical times exist at all temperatures and are shifted continuously by $\gamma$ and $\beta$.

Choosing the representative branch $\lambda_f=\pi/2$, we have
\begin{align}
N(t)&=\cos(\Omega t)+\gamma\chi\sin(\Omega t),
\nonumber\\
D(t)&=1+2\gamma^2\sin^2(\Omega t)+2\gamma\chi\sin(\Omega t)\cos(\Omega t).
\label{single-parameter-ND}
\end{align}
The corresponding critical times are
\begin{align}
t_n^{\mathrm F}(\beta)=\frac{\pi/2+\arctan(\gamma\chi)+n\pi}{\Omega},\qquad n\in\mathbb Z.
\label{finite-temperature-critical-times}
\end{align}
The product $\gamma\chi$ controls the relative weights of the cosine and sine terms in $N(t)$, and hence shifts the critical times.

Appendix~\ref{app:model-real-time-amplitude} derives these results and proves that $D(t)>0$ on the real-time axis. The normalization denominator thus neither generates nor shifts real-time critical points; it only modifies the amplitude and its temporal envelope. The normalized amplitude $\mathcal G_\rho(t)=N(t)/\sqrt{D(t)}$ changes sign when it crosses a Fisher-zero line.

\subsubsection{Temperature and Non-Hermiticity Effects}
\label{subsec:finite-temperature-critical-lines}

Each critical time decomposes into a fixed periodic part and a shift induced by temperature and non-Hermiticity. We define
\begin{align}
\Delta\tau(\beta,\gamma)
\equiv
\Omega t_n^{\mathrm F}-\left(\frac{\pi}{2}+n\pi\right)
=
\arctan\left[\gamma\tanh(\beta\Omega)\right].
\label{critical-time-shift}
\end{align}
This shift quantifies the correction to the Fisher-cancellation time caused by the post-quench non-Hermitian structure at fixed spectrum. Because the level spacing remains $2\Omega$, neither temperature nor $\gamma$ changes the fundamental period; they only tune the relative weights of the cosine and sine terms in $N(t)$ through $\gamma\chi$. Increasing $\gamma$ enhances the non-Hermitian contribution and delays destructive cancellation, while increasing temperature reduces $\chi$ and suppresses the shift. For $\gamma=0$, $\Delta\tau=0$, and the critical times reduce to their Hermitian values.

Importantly, the shift and the time-dependent normalization factor cannot be removed by any global similarity transformation, in sharp contrast to the metric-stationary quench studied in our previous work. This demonstrates that when the metric changes, the Loschmidt amplitude of the quasi-Hermitian system is fundamentally different from that of its Hermitian counterpart.

The zero- and infinite-temperature limits are
\begin{align}
\Omega t_n^{\mathrm F}
&\longrightarrow
\frac{\pi}{2}+\arctan\gamma+n\pi,
&&
T\longrightarrow0,
\nonumber\\
\Omega t_n^{\mathrm F}
&\longrightarrow
\frac{\pi}{2}+n\pi,
&&
T\longrightarrow\infty.
\label{critical-time-temperature-limits}
\end{align}
In the zero-temperature limit $\chi\to1$, the initial state approaches the ground-state projector and the non-Hermitian term in $N(t)$ has maximum weight, producing the largest shift $\arctan\gamma/\Omega$. As temperature increases, the thermal occupations of the two levels become more equal, weakening the response to the post-quench quasi-Hermitian structure; thus $\gamma\chi$ and the shift decrease continuously. In the infinite-temperature limit, $\rho_i\to I/2$ and $\chi\to0$, so $N(t)\to\cos(\Omega t)$ and the critical times become independent of $\gamma$. The strongest temperature dependence occurs when $k_BT\sim\Omega$, where the relative occupations are most sensitive.

\begin{figure}[t]
\centering
\includegraphics[width=\columnwidth]{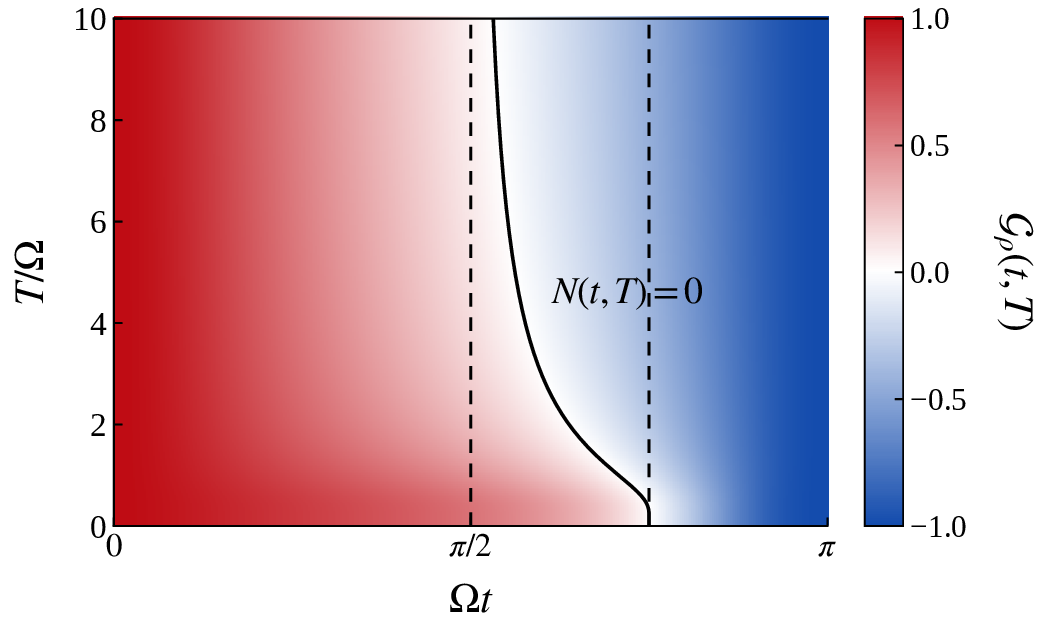}
\caption{Normalized mixed-state Loschmidt amplitude in the time--temperature plane for $\gamma=1$. The solid black curve is the Fisher-zero line, while the dotted and dashed lines denote the infinite- and zero-temperature limits, respectively.}
\label{fig:G-density}
\end{figure}

The derivatives of the critical times are
\begin{align}
\frac{\partial t_n^{\mathrm F}}{\partial\gamma}
&=
\frac{\chi}{\Omega\left(1+\gamma^2\chi^2\right)}>0,
\nonumber\\
\frac{\partial t_n^{\mathrm F}}{\partial\beta}
&=
\frac{\gamma\,\operatorname{sech}^2(\beta\Omega)}{1+\gamma^2\chi^2}\geq 0.
\label{critical-time-monotonicity}
\end{align}
At fixed temperature, the response to $\gamma$ weakens as $\gamma$ increases, indicating asymptotic saturation of the non-Hermitian shift. At fixed $\gamma$, $\partial t_n^{\mathrm F}/\partial\beta\to0$ in the low-temperature limit, where the initial state is already dominated by the ground state and further cooling leaves the Fisher-cancellation condition essentially unchanged.

Figure~\ref{fig:G-density} shows the normalized mixed-state Loschmidt amplitude in the time--temperature plane for $\gamma=1$. The first Fisher-zero line connects the zero-temperature limit $\Omega t_0^{\mathrm F}=3\pi/4$ to the infinite-temperature limit $\Omega t_0^{\mathrm F}=\pi/2$, and the remaining branches repeat periodically with spacing $\pi/\Omega$. The zero lines approach their limiting values in the low- and high-temperature regimes, while the principal temperature dependence occurs in the intermediate region. For this orthogonal quench, the normalized amplitude is real and satisfies $-1\leq\mathcal G_\rho(t)\leq1$; normalization modifies the amplitude profile but not the Fisher-zero locations.

\begin{figure}[t]
\centering
\includegraphics[width=0.9\columnwidth]{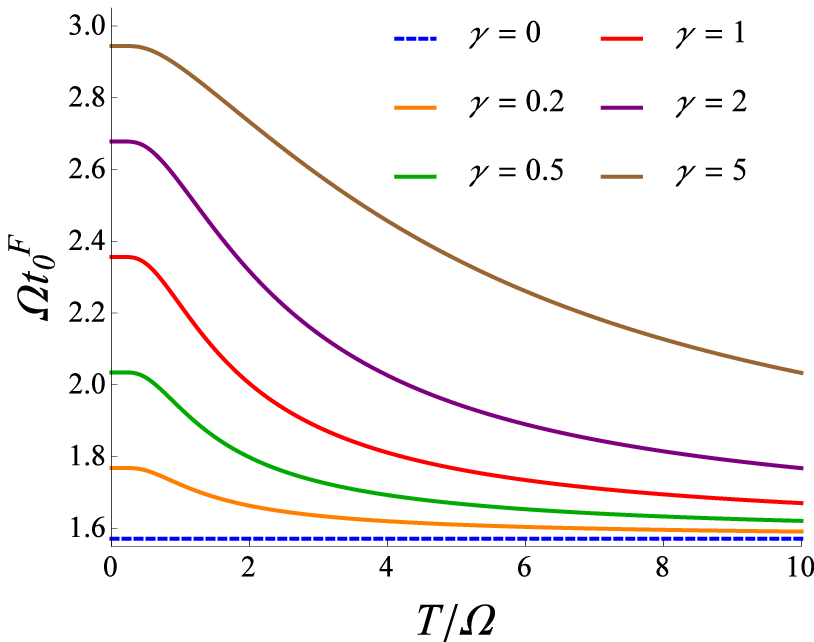}
\caption{Temperature dependence of the first real-time critical time $t_0^{\mathrm F}$ for $k_B=1$. Different curves correspond to different values of the non-Hermiticity parameter $\gamma$. The low- and high-temperature limits are $\Omega t_0^{\mathrm F}=\pi/2+\arctan\gamma$ and $\Omega t_0^{\mathrm F}=\pi/2$, respectively.}
\label{fig:tc-gamma}
\end{figure}

Figure~\ref{fig:tc-gamma} compares the temperature dependence of the first critical time $t_0^{\mathrm F}$ for different $\gamma$. At any finite temperature, $t_0^{\mathrm F}$ increases monotonically with $\gamma$, in agreement with Eq.~\eqref{critical-time-monotonicity}. The low-temperature plateaus are set by $\pi/2+\arctan\gamma$, showing that a strongly polarized initial state retains the non-Hermiticity-induced shifts. As temperature rises, the thermal polarization decreases, the curves approach one another, and their separation eventually disappears. Temperature therefore does not eliminate the finite-temperature DQPT; it shifts the critical time and suppresses the non-Hermiticity-induced shift in the high-temperature limit.

\section{Conclusion}
\label{sec:conclusion}

We have studied metric-changing quenches of quasi-Hermitian mixed states. Choosing the initial physical space as reference, we defined a normalized Loschmidt amplitude from purification overlap: the numerator is the dynamical overlap, while the denominator measures the non-quasi-unitary norm variation induced by metric mismatch. For real times the denominator is strictly positive, so criticality arises solely from destructive interference in the numerator. In complex time, denominator zeros generate branch singularities distinct from Fisher zeros.

For a general two-level system, we derived closed-form conditions for both classes of zeros and for inverse-square-root scaling near a second-order exceptional point. The gap-closing protocol separates divergent noncentral branches from finite central ones, while the isospectral protocol shows that finite-temperature real-time Fisher zeros require orthogonal quenches, with thermal polarization and non-Hermiticity shifting the critical times at fixed level spacing. Crucially, unlike the metric-stationary case, metric mismatch yields a time-dependent normalization and branch structure that cannot be removed by any global similarity transformation, confirming the genuinely non-Hermitian nature of these effects. The framework thus unifies thermal statistics, metric mismatch, and complex-time analytic structure in quasi-Hermitian mixed-state quenches.

\begin{acknowledgments}
H. G. was supported by the Innovation Program for Quantum Science and Technology--National Science and Technology Major Project (Grant No.~2021ZD0301904) and the National Natural Science Foundation of China (Grant No.~12447216). X.-Y. H. was supported by the National Natural Science Foundation of China (Grant No.~12405008).
\end{acknowledgments}

\appendix

\section{Positivity of the Normalization Function}
\label{app:positive-D}

We prove that $D(t)>0$ for all finite real $t$. Let the initial Hamiltonian $H_{\mathrm i}$ have the biorthogonal eigenbasis $\{|\psi_n^{(\mathrm i)}\rangle,|\phi_n^{(\mathrm i)}\rangle\}$ satisfying $H_{\mathrm i}|\psi_n^{(\mathrm i)}\rangle = E_n^{(\mathrm i)}|\psi_n^{(\mathrm i)}\rangle$ with $\langle\phi_m^{(\mathrm i)}|\psi_n^{(\mathrm i)}\rangle = \delta_{mn}$ and $\langle\phi_n^{(\mathrm i)}| = \langle\psi_n^{(\mathrm i)}|\eta_{\mathrm i}$. The initial thermal state is
\begin{align}
\rho_{\mathrm i}
=
\sum_n
p_n
|\psi_n^{(\mathrm i)}\rangle
\langle\phi_n^{(\mathrm i)}|,
\quad
p_n\geq0,
\quad
\sum_n p_n=1.
\end{align}
For real $t$,
\begin{align}
D(t)
&=
\operatorname{Tr}_{\eta_{\mathrm i}}
\left[
\rho_{\mathrm i}
\me^{\mi H_{\mathrm f}^{\ddagger_{\mathrm i}}t}
\me^{-\mi H_{\mathrm f}t}
\right]
\nonumber\\
&=
\sum_n p_n
\langle\psi_n^{(\mathrm i)}|
\eta_{\mathrm i}
\me^{\mi H_{\mathrm f}^{\ddagger_{\mathrm i}}t}
\me^{-\mi H_{\mathrm f}t}
|\psi_n^{(\mathrm i)}\rangle .
\end{align}
Using $H_{\mathrm f}^{\ddagger_{\mathrm i}}=\eta_{\mathrm i}^{-1}H_{\mathrm f}^{\dagger}\eta_{\mathrm i}$, we have $\eta_{\mathrm i}\me^{\mi H_{\mathrm f}^{\ddagger_{\mathrm i}}t}=\me^{\mi H_{\mathrm f}^{\dagger}t}\eta_{\mathrm i}$, so
\begin{align}
D(t)
&=
\sum_n p_n
\langle\psi_n^{(\mathrm i)}|
\me^{\mi H_{\mathrm f}^{\dagger}t}
\eta_{\mathrm i}
\me^{-\mi H_{\mathrm f}t}
|\psi_n^{(\mathrm i)}\rangle
\nonumber\\
&=
\sum_n p_n
\left\|
\me^{-\mi H_{\mathrm f}t}
|\psi_n^{(\mathrm i)}\rangle
\right\|_{\eta_{\mathrm i}}^2 .
\label{appendix-positive-D}
\end{align}
Because $\eta_{\mathrm i}>0$, every term in Eq.~\eqref{appendix-positive-D} is a nonnegative squared norm. In finite dimension, $\me^{-\mi H_{\mathrm f}t}$ is invertible for all finite real $t$, so no initial eigenstate with nonzero weight is mapped to zero; together with $\sum_n p_n=1$, this gives $D(t)>0$. The proof relies only on positivity of the metric and invertibility of the real-time evolution operator, not on $H_{\mathrm f}^{\ddagger_{\mathrm i}}=H_{\mathrm f}$.

Equivalently, since $W(t)=\me^{-\mi H_ft}W(0)$, $D(t)=\left(W(t),W(t)\right)_{\mathrm{HS}}^{\eta_i}$,
so $D(t)$ is the squared Hilbert--Schmidt norm of a nonzero vector. In complex time, the two exponential factors no longer form an adjoint pair, so $D(z)$ is not a squared norm and may have zeros.

\section{Two-Level Normalization Coefficients}
\label{app:two-level-coefficients}

Let $s,r=\pm1$ and define
\begin{align}
\widetilde P_s^{(\mathrm f)}
=
\left(P_s^{(\mathrm f)}\right)^{\ddagger_{\mathrm i}},
\quad
K_{sr}
=
\operatorname{Tr}_{\eta_{\mathrm i}}
\left[
\rho_{\mathrm i}
\widetilde P_s^{(\mathrm f)}
P_r^{(\mathrm f)}
\right].
\label{Ksr-definition}
\end{align}
Expanding the two evolution factors in $D(z)$,
\begin{align}
\me^{-\mi H_{\mathrm f}z}
&=
\me^{-\mi \bar E_{\mathrm f}z}
\sum_{r=\pm1}
\me^{-\mi r\Delta_{\mathrm f}z}
P_r^{(\mathrm f)},
\nonumber\\
\me^{\mi H_{\mathrm f}^{\ddagger_{\mathrm i}}z}
&=
\me^{\mi \bar E_{\mathrm f}z}
\sum_{s=\pm1}
\me^{\mi s\Delta_{\mathrm f}z}
\widetilde P_s^{(\mathrm f)} ,
\end{align}
we obtain
\begin{align}
D(z)
=
\sum_{s,r=\pm1}
K_{sr}\me^{\mi (s-r)\Delta_{\mathrm f}z}.
\end{align}
Thus the coefficients in Eq.~\eqref{general-D} are
\begin{align}
D_+=K_{+-},
\quad
D_-=K_{-+},
\quad
D_0=K_{++}+K_{--}.
\label{D-coefficients-Ksr}
\end{align}
The condition $D(0)=1$ gives $D_++D_-+D_0=1$.

Let $y=\me^{2\mi\Delta_{\mathrm f}z}$; then $D(z)=D_0+D_+y+D_-/y$, and zeros satisfy $D_+y^2+D_0y+D_-=0$. If $D_+=0$ or $D_-=0$, the quadratic reduces to a linear equation; since $y\neq0$ for finite $z$, a zero root does not correspond to a finite complex-time zero. If $D_+=D_-=0$, then $D_0=1$ and $D(z)=1$.

When $D_+D_-\neq0$ and $D_0^2-4D_+D_-=0$, the two roots coalesce at $y_0=-D_0/(2D_+)$. Since $\dif y/\dif z=2\mi\Delta_{\mathrm f}y$, this root is a second-order zero of $D(z)$; $\sqrt{D(z)}$ can then be chosen single-valued in its neighborhood, while the local behavior of $\mathcal G_\rho(z)$ also depends on the order of the zero of $N(z)$ there.

\section{Physical Metric in the Two-Level Model}
\label{app:model-metric}

We derive Eq.~\eqref{single-parameter-metric} and show how the measurement protocol fixes the metric.

Introduce the orthonormal frame
\begin{align}
\boldsymbol n_r
&=
(\sin\theta\cos\phi,\sin\theta\sin\phi,\cos\theta),
\nonumber\\
\boldsymbol n_\theta
&=
(\cos\theta\cos\phi,\cos\theta\sin\phi,-\sin\theta),
\nonumber\\
\boldsymbol n_\phi
&=
(-\sin\phi,\cos\phi,0),
\end{align}
with $\boldsymbol n_\mu\cdot\boldsymbol n_\nu=\delta_{\mu\nu}$ and $\boldsymbol n_\mu\times\boldsymbol n_\nu=\sum_{\rho}\epsilon_{\mu\nu\rho}\boldsymbol n_\rho$. Any $2\times2$ Hermitian matrix can be written as $\alpha I+\boldsymbol q\cdot\boldsymbol\sigma$ with real $\alpha,\boldsymbol q$. Expanding $\boldsymbol q=x\boldsymbol n_r+y\boldsymbol n_\theta+z\boldsymbol n_\phi$ gives
\begin{align}
\eta
=
\alpha I
+x\,\boldsymbol n_r\cdot\boldsymbol\sigma
+y\,\boldsymbol n_\theta\cdot\boldsymbol\sigma
+z\,\boldsymbol n_\phi\cdot\boldsymbol\sigma,
\label{general-metric-expansion}
\end{align}
with $\alpha,x,y,z\in\mathbb R$.

For the Hamiltonian
\begin{align}
H
=
\Omega
\left(
a\boldsymbol n_r+\mi b\boldsymbol n_\theta
\right)\cdot\boldsymbol\sigma,
\label{app-model-H-general}
\end{align}
where $\theta=\lambda$, $\phi=0$, $a=\sqrt{1+\gamma^2\sin^2\lambda}$, $b=\gamma\sin\lambda$, so $\boldsymbol n_\phi=(0,1,0)$ and $a^2-b^2=1$. Suppressing $\lambda$, we use $H^\dagger=\Omega(a\boldsymbol n_r-\mi b\boldsymbol n_\theta)\cdot\boldsymbol\sigma$ and $(\boldsymbol u\cdot\boldsymbol\sigma)(\boldsymbol v\cdot\boldsymbol\sigma)=(\boldsymbol u\cdot\boldsymbol v)I+\mi(\boldsymbol u\times\boldsymbol v)\cdot\boldsymbol\sigma$ to obtain
\begin{align}
H^\dagger\eta-\eta H
=
-2\mi \Omega b y\,I
+
2\mi \Omega
\left[
a y\boldsymbol n_\phi
-
\left(az+\alpha b\right)\boldsymbol n_\theta
\right]\cdot\boldsymbol\sigma.
\label{metric-condition-expanded}
\end{align}
The quasi-Hermiticity condition $H^\dagger\eta=\eta H$ thus requires $y=0$ and $az+\alpha b=0$. Defining $c=\alpha/a$ and $d=x$, we get
\begin{align}
\eta
=
c\left(aI-b\sigma_y\right)
+d\,\boldsymbol n_r\cdot\boldsymbol\sigma.
\label{general-compatible-metric-app}
\end{align}
Its eigenvalues are $\eta_\pm=ca\pm\sqrt{c^2b^2+d^2}$; with $a^2-b^2=1$, positivity requires $c>0$ and $|d|<c$. Since $d$ remains arbitrary, the Hamiltonian alone does not fix the metric.

To fix the metric, we include $\sigma_y$ as an additional observable, requiring $\sigma_y^\dagger\eta=\eta\sigma_y$. Since $[\sigma_y,\eta]=2\mi d\,\boldsymbol n_\theta\cdot\boldsymbol\sigma$, this forces $d=0$, so $\eta=c(aI-b\sigma_y)$. The determinant is $\det\eta=c^2(a^2-b^2)=c^2$; setting $\det\eta=1$ and $c>0$ gives $c=1$. Hence
\begin{align}
\eta
=
aI-b\sigma_y.
\end{align}

\section{Real-Time Amplitude in the Two-Level Model}
\label{app:model-real-time-amplitude}

With $\eta_i=I$ and $H_f^2=\Omega^2I$, the post-quench evolution operator is
\begin{align}
U_f(t)
=
\me^{-\mi H_ft}
=
\cos(\Omega t)I
-\mi \sin(\Omega t)\frac{H_f}{\Omega}.
\label{app:model-propagator}
\end{align}
From Eq.~\eqref{single-parameter-H},
\begin{align}
\frac{H_f}{\Omega}
&=
\left[
a(\lambda)\sin\lambda
+\mi \gamma\sin\lambda\cos\lambda
\right]\sigma_x
\nonumber\\
&\quad+
\left[
a(\lambda)\cos\lambda
-\mi \gamma\sin^2\lambda
\right]\sigma_z .
\label{app:model-H-components}
\end{align}
With $\rho_i=(I-\chi\sigma_z)/2$, we find
\begin{align}
N_\lambda(t)
=
\operatorname{Tr}_{\eta_i}
\left[
\rho_iU_f(t)
\right]
&=
\cos(\Omega t)
+\gamma\chi\sin^2\lambda\,\sin(\Omega t)
\nonumber\\
&\quad
+\mi \chi a(\lambda)\cos\lambda\,\sin(\Omega t).
\label{app:model-general-N}
\end{align}
At finite temperature, $\chi>0$. The condition $N_\lambda(t)=0$ is equivalent to
\begin{align}
a(\lambda)\cos\lambda\,\sin(\Omega t)
&=
0,
\nonumber\\
\cos(\Omega t)
+\gamma\chi\sin^2\lambda\,\sin(\Omega t)
&=
0.
\label{app:model-zero-conditions}
\end{align}
Since $a(\lambda)>0$ and $\sin(\Omega t)=0$ contradicts the second equation, we must have $\cos\lambda_f=0$; hence only orthogonal quenches $\lambda_f=\pi/2\pmod{\pi}$ can yield finite-temperature DQPTs.

For the representative case $\lambda_f=\pi/2$, define $c=\cos(\Omega t)$, $s=\sin(\Omega t)$. Then
\begin{align}
U_f(t)
&=
cI
-\mi \sqrt{1+\gamma^2}\,s\,\sigma_x
-\gamma s\,\sigma_z,
\nonumber\\
U_f^{\ddagger_i}(t)
&=
cI
+\mi \sqrt{1+\gamma^2}\,s\,\sigma_x
-\gamma s\,\sigma_z.
\label{app:model-orthogonal-propagators}
\end{align}
The Pauli algebra gives
\begin{align}
U_f^{\ddagger_i}(t)U_f(t)
&=
\left(
1+2\gamma^2s^2
\right)I
-2\gamma sc\,\sigma_z
\nonumber\\
&\quad
-2\gamma\sqrt{1+\gamma^2}\,s^2\sigma_y.
\label{app:model-propagator-product}
\end{align}
Using $\operatorname{Tr}_{\eta_i}(\rho_i\sigma_z)=-\chi$ and $\operatorname{Tr}_{\eta_i}(\rho_i\sigma_y)=0$, we obtain
\begin{align}
N(t)
&=
c+\gamma\chi s,
\nonumber\\
D(t)
&=
\operatorname{Tr}_{\eta_i}
\left[
\rho_iU_f^{\ddagger_i}(t)U_f(t)
\right]
=
1+2\gamma^2s^2+2\gamma\chi sc.
\label{app:model-ND}
\end{align}
Rewriting $D(t)$ in double angle,
\begin{align}
D(t)
=
1+\gamma^2
-\gamma^2\cos(2\Omega t)
+\gamma\chi\sin(2\Omega t),
\label{app:model-D-double-angle}
\end{align}
and using $\min_x(A\cos x+B\sin x)=-\sqrt{A^2+B^2}$ with $A=-\gamma^2$, $B=\gamma\chi$, gives
\begin{align}
D_{\min}
=
1+\gamma^2-\sqrt{\gamma^4+\gamma^2\chi^2}.
\end{align}
Since $0\leq\chi\leq1$,
\begin{align}
(1+\gamma^2)^2
-\left(
\gamma^4+\gamma^2\chi^2
\right)
=
1+\gamma^2(2-\chi^2)>0,
\end{align}
so $D_{\min}>0$.

\bibliography{Review}

\end{document}